\documentclass[aps,prl,twocolumn,floats,showpacs]{revtex4}

\usepackage{amsmath,amssymb,graphicx}
\usepackage{epsfig}
\usepackage{graphicx}
\usepackage{enumerate}
\newcommand{\beq}{\begin{equation}}
\newcommand{\eeq}{\end{equation}}

\newcommand{\bea}{\begin{eqnarray}}
\newcommand{\eea}{\end{eqnarray}}

\begin{document}

\title{Zeno and anti-Zeno dynamics in spin-bath models}

\author{Dvira Segal, David R. Reichman}
\affiliation{Department of Chemistry, Columbia University, 3000 Broadway, New York, NY 10027}

\begin{abstract}
We investigate the quantum Zeno and anti-Zeno effects in spin bath
models: the spin-boson model and a spin-fermion model. We show that the
Zeno-anti-Zeno transition is critically
controlled by the system-bath coupling parameter, the
same parameter that determines spin decoherence rate.
We also discuss the crossover in a biased system,  at high temperatures,
and for a nonequilibrium spin-fermion system,  manifesting the
counteracting roles of electrical bias, temperature,
and magnetic field on the spin decoherence rate.

\end{abstract}

\pacs{
03.65.Xp, 
03.65.Yz, 
05.30.-d, 
31.70.Dk  
}
\maketitle

The quantum Zeno effect (QZE) describes the behavior of a quantum system when
frequent short time measurements inhibits decay \cite{QZE,Ketterle}.
In some cases, however, the anti-Zeno effect (AZE), namely
an enhancement of the decay due to frequent measurements,
is observed \cite{AZE, Kurizky}.
The QZE can be easily obtained for an oscillating (reversible)
quantum system. When the system is unstable, the situation is more involved, and
one can obtain both the QZE and AZE, depending on the interaction Hamiltonian
\cite{Kurizky}, as well as  the measurement interval \cite{Facchi}.
A crossover from QZE to AZE behavior has been
observed in an unstable trapped cold atomic system via a tuning of
the measurement frequency \cite{Raizen}.
Recently, Maniscalco {\it et al.} \cite{QBM} have theoretically investigated the Zeno-anti-Zeno crossover in a model of a damped quantum harmonic oscillator.
These authors  demonstrated the crucial role played by the short
time behavior of the environmentally induced decoherence.

In this letter we focus on spin-bath models,
paradigms of quantum dissipative systems,
and analyze the conditions for the occurrence of the Zeno and
anti-Zeno effects. We show that the
crossover between the two processes is critically controlled by the
dimensionless coupling strength, as well as the temperature and
the energy bias between the spin states.
We also analyze the Zeno dynamics
for out-of-equilibrium, electrically biased, situations.
Our main result is that the same parameters
which determine the extent of quantum coherence for the transient
population variable,
also control the Zeno-anti-Zeno transition.
Therefore, the nature of spin decoherence can be predicted from the Zeno dynamics.

The models of interest here are
the spin-boson model and a nonequilibrium steady state spin-fermion model, 
which is a simplified variant of the Kondo model.
The spin-boson model, describing a two-level system coupled to a
bath of harmonic oscillators, is one of the most important models
for elucidating the effect of environmentally controlled
dissipation in quantum mechanics \cite{Legget}.
The Kondo model, possibly the simplest model of a magnetic impurity coupled to an
environment, describes the coupling of a
magnetic atom to the conduction band electrons \cite{Kondo}.
This system has recently regained enormous interest due to 
significant experimental progress  in mesoscopic physics \cite{Kondo-Rev}.
For both models the prototype Hamiltonian includes three contributions
\beq
H=H_S+H_B+H_{SB}.
\label{eq:Htotal}
\eeq
The spin system includes a two level system (TLS) with
a bare tunneling amplitude $\Delta$ and a level splitting $B$,
\beq
H_S=\frac{B}{2}\sigma_z+\frac{\Delta}{2} \sigma_x.
\eeq
In what follows we refer to the bias $B$ as a magnetic field in order to distinguish it
from potential bias in a  nonequilibrium system.
In the bosonic case the thermal bath includes a set of independent
harmonic oscillators, and the system-bath interaction is bilinear 
%
\bea H_B^{(b)}&=&\sum_{j}\epsilon_j b_j^{\dagger}b_j,
\nonumber\\
H_{SB}^{(b)}&=&\sum_{j}\frac{\lambda_j}{2}(b_j^{\dagger}+b_j)\sigma_z.
\label{eq:Hboson} \eea
$b_j^{\dagger}, b_j$ are bosonic creation and annihilation operators,
respectively. 
For a fermionic system we employ the model
%
\bea
H_B^{(f)}&=&\sum_{k}\epsilon_k a_{k,n}^{\dagger}a_{k,n},
\nonumber\\
H_{SB}^{(f)}&=&\sum_{k,k',n,n'}\frac{V_{k,n;k',n'}}{2}a_{k,n}^{\dagger}a_{k',n'}\sigma_z.
\label{eq:Hfermion} \eea
Here $a_k^{\dagger}, a_k$ are fermionic creation and annihilation operators, and
the spin interacts with $n$ reservoirs.
We also define two auxiliary Hamiltonians $H_{\pm}$ as
\beq H_{\pm}=\pm \frac{B}{2}+H_{SB}(\sigma_z=\pm)+H_B.
\label{eq:Hpm} \eeq
Note that the Hamiltonian (\ref{eq:Htotal}) does not include explicitly the measuring device.
While in Refs. \cite{Prezhdo} (boson bath) and \cite{Gurvitz} (fermion bath)
the reservoirs serve as continuous detectors, here they are part of the dynamical system
under measurement.

Before studying the Zeno dynamics of the dissipative systems,
we briefly review the results for an isolated TLS.
For zero magnetic field, the probability to remain in the initial
prepared state is given by
%
$W(t)= \cos^2 (\Delta t/2).$
%
The short-time dynamics ($\Delta t\ll 1$ ) can be approximated by
%
$W(t)\sim 1-(\Delta/2)^2t^2$.
%
If measurements are performed at regular intervals $\tau$,
the survival probability at time $t=n\tau$ becomes
%
$W(t) \sim   1-n(\Delta/2)^2\tau^2
\sim  \exp(-\Delta^2\tau t/4).$ 
%
At short times an effective relaxation rate is identified as
\beq
\gamma_0(\tau)=(\Delta/2)^2\tau.
\label{eq:gamma0}
\eeq
As $\tau$ goes to zero, decay is inhibited,
and the dynamics is frozen. This is the quantum Zeno effect.


We consider next the influence of the environment on this behavior.
Within perturbation theory the
dynamics of a TLS coupled to a general heat bath is given as  a power series of
$\Delta$ terms,
$W(t)=\sum_{n=0}^{\infty} (\frac{\Delta}{2})^{2n}\Phi_n(t)$ \cite{Chang}.
We have numerically verified that for $\omega_c \tau \lesssim 20$,
($\omega_c$ is the reservoir cutoff frequency)
the series can be approximated by the first two terms
\bea
W(\tau)&\sim&
1-2 \Re \left(  \frac{\Delta}{2} \right)^2\int_0^{\tau}dt_1\int_0^{t_1}dt_2
K(t_2),
\nonumber\\
K(t)&=& \langle e^{-iH_+t}e^{iH_-t}\rangle,
\label{eq:K}
\eea
even for strong system-bath coupling \cite{App},
provided that $\Delta \tau<1$ \cite{Series}.
Here $\Re$ denotes the Real part,
the Hamiltonians $H_{\pm}$ are defined in Eq. (\ref{eq:Hpm}), and the trace is
done over the bath degrees of freedom, irrespective of the statistics.
Similarly to the isolated case,
we can identify an effective decay rate at short times
\beq
\gamma(\tau)=  \frac{\Delta^2} {2\tau} \Re \int_0^{\tau}dt_1 \int_0^{t_1}K(t')dt'.
\label{eq:gamma}
\eeq
In what follows we disregard the multiplicative factor $(\Delta/2)^2$.
The basic question to be addressed in this letter is how does
this relaxation rate depend on the system bath coupling.
It is clear that when the system is decoupled from the environment,
 Eq. (\ref{eq:gamma}) reproduces the result of the isolated
system, Eq. (\ref{eq:gamma0}). For a dissipative system the
central object of our calculation is therefore the correlation function $K(t)$
defined in Eq. (\ref{eq:K}).
For the bosonic system (\ref{eq:Hboson}), a standard calculation yields \cite{Mahan}
\bea K_b(t)&=&e^{-i E_s t} \exp
\bigg\{-\frac{1}{\pi}\int_0^{\infty}d\omega \frac{J(\omega)}{\omega^2}
\nonumber\\
&\times&
\left[(n_{\omega}+1)\left( 1-e^{-i\omega t}\right)
+n_{\omega}\left( 1-e^{i\omega t}\right) \right]
\bigg\}.
\label{eq:Kbg}
\eea
Here $n_{\omega}=[e^{\beta \omega}-1]^{-1}$ is the Bose-Einstein
distribution function, $\beta=1/T$ is the inverse temperature, and
$E_s=B+\int \frac{J(\omega)}{\pi \omega}  d\omega$ is the polaron
shift. The spectral density $J(\omega)$ includes the information
about the system-bath interaction
%
$J(\omega)=\pi\sum_{j}\lambda_j^2\delta(\omega-\omega_j)$.
%
For an ohmic model, $J(\omega)=2 \pi \xi \omega
e^{-\omega/\omega_c}$, Eq. (\ref{eq:Kbg}) becomes \cite{Aslangul}
\beq K_b(t)= \left(\frac{1}{1+i\omega_ct}\right)^{2\xi} \left[
\frac{\left( \frac{\pi t}{\beta}\right) } {\sinh\left( \frac{\pi
t}{\beta}\right)} \right]^{2\xi} \times e^{-i E_s t}
\label{eq:Kb}
\eeq
where $\xi$ is a system-bath dimensionless coupling parameter \cite{Legget}.
At zero temperature and for zero magnetic field, disregarding the
energy shift, the effective decay rate (\ref{eq:gamma}) is given
by ($\xi\neq \frac{1}{2}, 1$)
\bea
\gamma(\tau)&=&\frac{ 1-(1+\omega_c^2\tau^2)^{1-\xi}\cos[2(1-\xi)\rm{atan}(\omega_c\tau)]}
{\tau\omega_c^2(1-\xi)(1-2\xi)}.
\nonumber\\
\label{eq:T0a}
\eea
%
\begin{figure}
{\hbox{\epsfxsize=80mm \epsffile{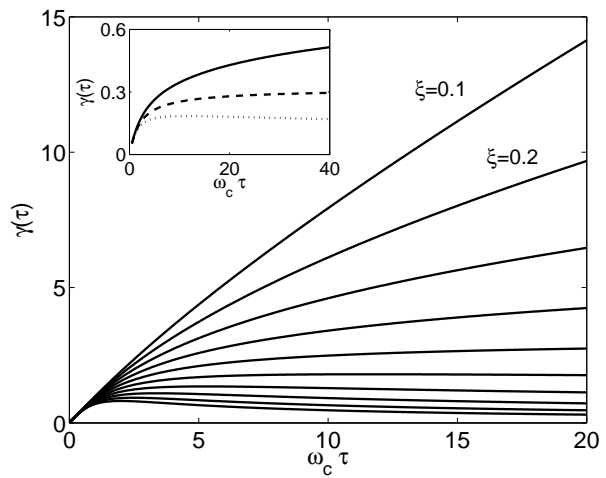}}}
\caption{The effective decay rate for a spin coupled to a bosonic bath
at zero temperature for 
$\omega_c$=1, ${E}_s=0$, $\xi$=0.1, 0.2, 0.3...1,
top to bottom.
The appearance of a maximum point in $\gamma(\tau)$ around $\omega_c \tau=2$ 
for $\xi>1/2$ marks the transition from QZE to AZE. 
The inset depicts the rate for $\xi$=0.4 (full), 0.5 (dashed), 0.6 (dotted),  $\omega_c$=10.}
 \label{figb1}
\end{figure}
%
The limiting behavior of this expression is
\beq
\gamma(\tau) \propto
\begin{cases}
\tau & \omega_c\tau<1\\
\tau^{1-2\xi} & \omega_c \tau>1.
\end{cases}
\label{eq:T0}
\eeq
Therefore, while at very short times ($\omega_c \tau<1$) the
system always shows the QZE, irrespective of the system-bath
coupling, at intermediate times  $1/\omega_c< \tau<1/ \Delta$ the
qualitative behavior of the decay rate crucially depends  on the
dimensionless coupling coefficient. The decay is inhibited (QZE)
for $\xi<\frac{1}{2}$, and accelerated (AZE) for
$\xi>\frac{1}{2}$. At $\xi=\frac{1}{2}$ the decay rate does no
depend on the measurement interval,
%
$\gamma(\tau)\xrightarrow{\frac{\log(\omega_c \tau)}{\omega_c\tau}<1}\frac{\pi \Delta^2}{4\omega_c}$.
%
Correspondingly, at zero temperature an unbiased spin oscillates
coherently at $\xi=0$, shows damped harmonic oscillation for
$0<\xi<\frac{1}{2}$, and decays incoherently at strong
dissipation, $\frac{1}{2}<\xi<1$. At the crossover value
$\xi=\frac{1}{2}$ the system can be mapped onto the Toulouse
problem, describing an impurity coupled to a Fermi bath with a
constant density of states. This leads to a purely Markovian
exponential decay which is unaffected by measurements
\cite{Toulouse}. We therefore find that the dimensionless coupling
constant $\xi$ which controls the extent of spin decoherence,
determines the occurrence of the QZE and the AZE.

Besides the occupation probability $W(t)$,
the symmetrized  equilibrium correlation function $C(t)$ is also of interest.
It is unclear whether the coherent-incoherent transition of this quantity
occurs at $\xi=\frac{1}{2}$ \cite{Mak}, 
or below, at $\xi=\frac{1}{3}$ \cite{Saleur}.
Yet, since the short-time dynamics of the two-spin correlation function
$C(t)$ is controlled by the same function $K(t)$ 
[Eq. (\ref{eq:K})] \cite{Weiss},
its Zeno-anti-Zeno behavior {\it exactly corresponds with $W(t)$},
and thus rigorously undergoes a Zeno-anti-Zeno transition at $\xi=\frac{1}{2}$. 



\begin{figure}
 {\hbox{\epsfxsize=75mm \epsffile{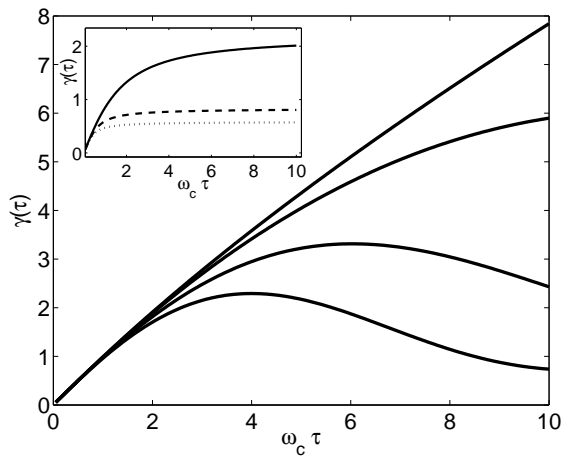}}}
 \caption{
The Zeno-anti-Zeno transition of a spin-boson model under a magnetic field,
 $T=0 K$, $\omega_c$=1, $\xi=0.1$, $B$=0, 0.2, 0.4, 0.6, top to bottom.
Inset: Temperature effect, 
$\beta \omega_c =0.5$, $B$=0,
$\xi=0.1$ (full),  $\xi=0.4$ (dashed),  $\xi=0.7$ (dotted).}
 \label{figb2}
\end{figure}


Fig. \ref{figb1} depicts the decay rate $\gamma(\tau)$ for
$\xi=0.1-1$, manifesting the crossover from the QZE to the 
AZE at $\xi=\frac{1}{2}$ (inset).
The transition between the Zeno and anti-Zeno regimes can be manipulated
by applying a finite magnetic field and by employing a finite temperature bath.
Fig. \ref{figb2} shows that at weak coupling, a biased
system $B/\omega_c\sim 1$ tends towards a pronounced anti-Zeno behavior, in  
contrast to the zero magnetic field case.
The influence  of finite temperatures is radically different (inset):
It drives the system into the Zeno regime,
for all coupling strengths. This can be deducted from Eq. (\ref{eq:Kbg}):
For $\beta \omega_c <1$,
$K(t)\sim e^{-2\pi\xi t/\beta}$, and  the resulting decay rate is proportional
to $\tau$ at short times, saturating at $\omega_c\tau>1$.

Our results can be elucidated by recasting the decay rate as
a convolution of two memory functions,
\bea
\gamma(\tau)=2 \left(\frac{\Delta}{2}\right)^2 \int_0^{\infty} d\omega K(\omega) F_{\tau}(\omega),
\label{eq:overlap}
\eea
where the measurement function is given by
$F_{\tau}(\omega)=\frac{\tau}{2\pi} {\rm sinc}^2\left[ \frac{(\omega- E_s)\tau}{2}\right]$,
%
and $K(\omega)=\Re\int_0^{\infty}e^{i\omega t}K(t) dt$,
can be interpreted as the reservoir coupling spectrum \cite{Kurizky}.
Note that $K(t)$ is redefined here without the energy shift.
The short-time behavior is therefore determined by the overlap of these two memory functions.
For $| E_s-\omega_m|\gg 1/\tau$, with $\omega_m$ as the central frequency of $K(\omega)$,
AZE takes place, while for a bath with narrow coupling spectrum and
for $| E_s-\omega_m|\ll 1/\tau$ the QZE occurs  \cite{Kurizky}.

\begin{figure}
 {\hbox{\epsfxsize=72mm \epsffile{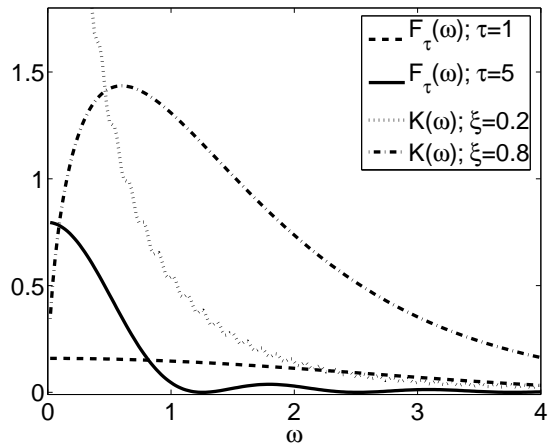}}}
 \caption{The QZE and the AZE,
as observed through the spectral functions $K(\omega)$
and $F_{\tau}(\omega)$ for $T=0$, $\omega_c$=1, $ E_s=0$
(see Eq. (\ref{eq:overlap})  for definitions).
Increasing $\xi$, $K(\omega)$ shifts towards 
higher frequencies overlapping poorly with $F_{\tau=5}(\omega)$, 
and strongly with $F_{\tau=1}(\omega)$,
which leads to the AZE.
}
\label{figb4}
\end{figure}

Fig. \ref{figb4} displays  $K(\omega)$ at two values:  $\xi$=0.2, 0.8.  
We also show the measurement function
$F_{\tau}(\omega)$ for $\omega_c \tau$=1, 5.
We find that at weak coupling, the overlap between these two functions decreases upon shortening $\tau$,
leading to the QZE.
In contrast, at strong coupling, since the central frequencies
 of $F_{\tau}(\omega)$ and $K(\omega)$ are detuned,
increasing the width of $F_{\tau}(\omega)$ enhances the overlap between the functions,
therefore the decay rate, leading to the AZE.
This is clearly seen mathematically: At zero temperature
$K(\omega)=\frac{1}{\omega_c\Gamma(2\xi)} \left(\frac{\omega}{\omega_c}\right)^{2\xi-1} e^{-\omega/\omega_c}$, with
$\omega_m\sim(2\xi-1)\omega_c$.
This argument also explains the AZE observed at weak
coupling for finite magnetic fields.
Since $F_{\tau}(\omega)$ is centered around $B$, and $\omega_m\sim 0$
for $\xi \le 0.5$,
the AZE is expected to prevail for $B \tau > 1$, as seen
in Fig. \ref{figb2}.


We turn now to the fermionic bath. For a single reservoir at zero
temperature, and for times $D\tau>1$, the fermionic
correlation function can be calculated in the context of the
Fermi-edge singularity problem \cite{Nozieres},
\beq K_f(t)\sim \frac{e^{-iB
t}}{\left(1+iDt\right)^{\delta^2/\pi^2}}. \label{eq:Kfe} \eeq
Here $\delta=\rm{atan}(\pi \rho V)$,
$\rho$ is the density of states, $D$ is the bandwidth, the equivalent
of the cutoff frequency $\omega_c$ in the bosonic case,  and we used a
constant coupling model $V_{k,k'}=V$.
We have disregarded the energy shift coming from the diagonal
coupling $V_{k,k}$, as it can always be  accommodated into the external magnetic field.

Comparing Eq. (\ref{eq:Kfe}) to the bosonic expression
(\ref{eq:Kb}), leads to the conclusion that an unbiased spin
coupled to a fermionic bath can only manifest the QZE: Since the
phase shift  cannot exceed  $\pi/2$, the exponent in
(\ref{eq:Kfe}) is always $\leq1/4$, while according to Eq.
(\ref{eq:T0}), the AZE takes place only for larger values which
are prohibited in this case. A spin coupled to more than a single
lead may attain a larger exponent, which can lead to the AZE
\cite{fermion}.

The nonequilibrium spin-fermion system, where the spin couples to
{\it two} fermionic baths ($V_{k,n;k'n'}=V$, $n=1,2$) with
chemical potentials  shifted by $\delta \mu$, is much more
interesting and involved. In this case we numerically calculate
the correlation function $K_{f}(t)$, since its analytic form is
not known at short times $Dt\leq1$ and for arbitrary voltages
\cite{Ng}. This is done by expressing the zero temperature many
body average as a determinant of the single particle correlation
functions \cite{Mahan}
\bea
K_f(t)&=&\langle e^{-iH_+t} e^{iH_-t} \rangle
= \det \left[ \phi_{k,k'}(t)  \right]_{k,k'<k_f},
\nonumber\\
\phi_{k,k'}(t)&=&\langle k| e^{-ih_+t}e^{ih_-t} |k' \rangle.
\eea
Here $H_{\pm}=\sum h_{\pm}$ are the single particle Hamiltonians,
$|k\rangle$ are the single particle eigenstates of $H_B^{(f)}$ and the
determinant is evaluated over the occupied states.

For a short-time evolution, it is satisfactory to model
the fermionic reservoirs  using 200 states per bath, where bias is applied by
depopulating one of the reservoirs with respect to the other.
The decay rate $\gamma(\tau)$, (Eq. (\ref{eq:gamma})), is presented in
the main plot  for three situations:
In the absence of both magnetic field and electric bias (full), including a finite magnetic
field (dashed), and  under an additional potential bias (dotted).
We find that the potential bias in the fermionic  system
plays the role of the temperature in the bosonic environment \cite{Ng},
driving the anti-Zeno behavior into a Zeno dynamics.

Summarizing,
we find that the same coupling parameter that monitors
spin decoherence and relaxation,  determines the Zeno behavior.
In addition, while finite temperature and electric bias eliminate the AZE,
applying a finite magnetic field can revive the effect.
The relationship between the Zeno effect and spin dynamics
also implies on the feasible
{\it control} over the environment induced spin decoherence 
utilizing the Zeno effect, which is crucial for quantum computing applications \cite{Hosten}.

The transition from Zeno to anti-Zeno dynamics can be controlled
by modifying the environmental parameters such as the
spectral density, temperature,  electrical bias,
and by changing the system-bath interaction, as well as the spin parameters.
Trapped ions in an optical lattice is a highly versatile system possibly capable
of showing the Zeno-anti-Zeno crossover.
In recent years it has become feasible to trap chains of atoms, to couple them in a controlled way
to the oscillatory "phonon" modes of the chain, and to probe them by a laser field \cite{Bloch}.
%
Another possible setup is an atomic Bose-Einstein condensate
interacting with a laser field \cite{Ketterle}.
This system is predicted to give rise to composite
quasiparticles: local impurities dresses by (virtual) phonons  \cite{Davidson}.
The spin-fermion model could  be realized in a semiconductor
microstructure consisting of two coupled quantum dots, simulating a double-well potential,
interacting with a current carrying quantum point contact (QPC) \cite{Hack}.
Measurements of the spin state can be done either continuously with an additional QPC,
serving this time as a detector, or using laser radiation, directly detecting the
population in each of the wells.

\begin{figure}
 {\hbox{\epsfxsize=80mm \epsffile{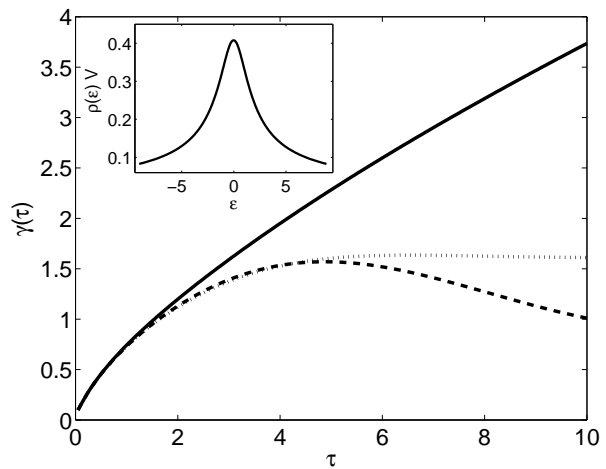}}}
 \caption{The effective decay rate for a spin coupled to two fermionic baths demonstrating the counteracting roles of the magnetic field and
 electrical bias.
$B$=0, $\delta\mu$=0 (full); $B$=0.5, $\delta\mu$=0 (dashed); $B$=0.5, $\delta\mu$=0.2 (dotted).
The inset shows the Lorentzian shaped density of states.
}
 \label{figf1}
\end{figure}

\acknowledgments
This work was supported by NSF (NIRT)-0210426.



\begin{thebibliography}{99}
\bibitem{App}
We define "strong system-bath coupling" as  $\xi>0.5$
(bosons) and  $\pi \rho V\gtrsim 1$ (fermions).
These parameters are defined after Eqs. (\ref{eq:Kb}) and  (\ref{eq:Kfe})
respectively.
\bibitem{Series}
In the discussion below we always assume $\Delta \tau<1$.
\end{thebibliography}
\end{document}